\documentclass[preprint,12pt,longnamesfirst]{aastex}
\usepackage{emulateapj5}

\newcommand{\kms}{\hbox{ km~s$^{-1}$}}    

\slugcomment{The Astronomical Journal, 2004 November Issue}

\shortauthors{Goldman et al.}
\shorttitle{Physical Structure of NGC\,1360}

\begin{document}
\title{Physical Structure of Planetary Nebulae. III. 
The Large and Evolved NGC\,1360}


\author{Daniel B.\ Goldman, Mart\'{\i}n A.\ Guerrero\altaffilmark{1}, 
You-Hua Chu\altaffilmark{2}, and  Robert A.\ Gruendl}
\affil{Astronomy Department, University of Illinois at 
 Urbana-Champaign, 1002 West Green Street, Urbana, IL 61801}
\email{dgoldman@uiuc.edu, mar@astro.uiuc.edu, chu@astro.uiuc.edu,
  gruendl@astro.uiuc.edu}

\altaffiltext{1}{
Current address: Instituto de Astrof\'{\i}sica de Andaluc\'{\i}a,
Consejo Superior de Investigaciones Cient\'{\i}ficas (CSIC),
Apartado Correos 3004, E-18080, Granada, Spain.}
\altaffiltext{2}{
Visiting Astronomer, Cerro Tololo Inter-American Observatory,
National Optical Astronomy Observatories, operated by the
Association of Universities for Research in Astronomy, Inc.\ (AURA)
under a cooperative agreement with the National Science Foundation.}

\begin{abstract}

NGC\,1360 is a large planetary nebula (PN) without an obvious
shell morphology.
We have analyzed H$\alpha$ images and high-dispersion echelle 
spectra of NGC\,1360 in order to construct spatio-kinematic
models and to determine its density distribution.
The best-fit model indicates that NGC\,1360 is a prolate 
ellipsoidal shell whose major axis is twice as long as its
minor axis and is tilted by 60\arcdeg\ with respect to the 
line of sight.  
The large kinematic age of the shell, $\sim$10,000 yr, and
the low density of the nebula, $\le$130 H-atom cm$^{-3}$, 
imply that NGC\,1360 is an evolved PN and has begun to 
merge with the interstellar medium.  
The observed morphology and surface brightness profiles of 
NGC\,1360 can be described well as a thick shell with a
Gaussian radial density profile without a sharp inner
edge, indicating a lack of on-going compression by a 
fast stellar wind.
The FLIERs observed in NGC\,1360 near the end of its major
axis expand faster and are younger than the nebular shell.

\end{abstract}



\section{Introduction}

Planetary nebulae (PNe) exhibit a wide range of morphologies.
While the overall shape of a PN may be round, elliptical, or
bipolar, most are limb-brightened, indicating a lower 
density in the interior, or the existence of a central cavity.
It has been suggested that PNe are formed by dynamical 
interactions between the current fast stellar wind and previous 
slow wind \citep{KPF78}.
Hydrodynamic calculations of interacting-stellar-wind models have
indeed been successful in reproducing the hollow-shell structure 
and overall shapes of PNe by varying the pole-to-equator density 
gradient in the Asymptotic Giant Branch (AGB) progenitor's slow 
wind \citep[e.g.,][]{FM94}.
A shortcoming of the hydrodynamic modeling is that often several 
sets of parameters based on differing geometries are capable of
reproducing the same morphology.
Observations of spatially resolved kinematics of a PN can help 
differentiate among models.

We have been studying the physical structure of PNe using both 
imaging and high-dispersion spectroscopic observations, in order 
to understand their formation and evolution.
Thus far, we have analyzed the Owl Nebula (NGC\,3587) in Paper I
\citep{Getal03} and NGC\,7662 in Paper II \citep{Getal04}.
The Owl Nebula is a triple-shell PN; its ``owl face'' morphology is 
caused by bipolar cavities excavated by a fast wind that ceased
some time ago, and its kinematics indicate that the nebular material
is backfilling these cavities. 
NGC\,7662 is also a triple-shell PN; its bright main nebula consists 
of two elliptical shells with simple structures in H$\alpha$ and 
[\ion{O}{3}], although the outer rim of the main nebula is 
interspersed with [\ion{N}{2}]-bright knots.
These two PNe were chosen for our study because their morphologies
have been successfully reproduced by interacting-stellar-wind models.

There exist PNe that do not possess morphological features to 
suggest the presence of wind-wind interactions. 
NGC\,1360 is such a PN.  
Its surface brightness does not dip deeply at the center or 
rise steeply at the limb to indicate a hollow shell structure.
The lack of shell morphology in NGC\,1360 makes it particularly
interesting.
Therefore, we have chosen this PN for a spatio-kinematic analysis
in order to determine its physical structure and to investigate its
evolutionary status.
This paper reports our observations and analysis of NGC\,1360.

\section{Observations}

\subsection{Direct Images}

Narrow-band H$\alpha$ images of NGC\,1360 were obtained on 1999 
October 8--9 with the 1~m telescope at Mount Laguna Observatory (MLO).
The central wavelength and FWHM of the H$\alpha$ filter are 6563 \AA\ 
and 17 \AA, respectively.  
The detector was a Loral 2K CCD with a pixel size of 15 $\mu$m, 
corresponding to an image scale of $0\farcs4$ pixel$^{-1}$ and 
providing a field of view of 13\farcm3$\times$13\farcm3.  
The angular resolution was $\sim2\farcs8$, as determined from the 
FWHM of field stars in the images.  
A total of 10 images, each with an exposure time of 600 s, were 
combined to form the final image.  
This image is shown in Figure~1.  

We have also retrieved archival {\em Hubble Space Telescope} images of
NGC\,1360 taken with the Wide Field and Planetary Camera 2 (WFPC2)
and the F656N (H$\alpha$) and F658N ([\ion{N}{2}]) filters.  
These observations were obtained on 1994 February 15, during the 
``pre-cooldown" period when the WFPC2 was operating at a temperature 
of $-$77~{\arcdeg}C (rather than the optimal $-$88~{\arcdeg}C at which it normally 
operates).  
This higher operating temperature produced a substantially greater 
number of warm pixels and a higher dark current.
To process the observations we first flagged and interpolated 
the warm pixels using the hot pixel lists available at 
STScI\footnote{
http://www.stsci.edu/instruments/wfpc2/wfpc2$\_$hotpix.html},
and then proceeded with the standard routines in the STSDAS 
package within IRAF.
These images are not shown in this paper because the F658N
image did not detect any nebulosity and the
F656N image does not show any small-scale features unresolved 
in our ground-based H$\alpha$ image in Figure 1.
The F656N image has only been used to flux-calibrate our 
ground-based H$\alpha$ image.
The total H$\alpha$ flux from NGC\,1360 is $\sim$3.1$\times10^{-10}$ 
ergs cm$^{-2}$ s$^{-1}$.

\subsection{High-Dispersion Spectroscopic Observations}

High-dispersion spectroscopic observations of NGC\,1360 were made
using the echelle spectrograph on the 4~m telescope at Cerro 
Tololo Inter-American Observatory (CTIO) on 2001 December 25. 
The 79 line~mm$^{-1}$ echelle grating and the long-focus red camera 
were used to achieve a reciprocal dispersion of 3.4 \AA~mm$^{-1}$. 
The spectrograph was used in the single-order, long-slit mode to
allow the coverage of the H$\alpha$ and [N~{\sc ii}] 
$\lambda\lambda$6548,6584 lines along a slit length of $\sim$3$'$.
The data were recorded with the SITe 2K No.~6 CCD with a pixel 
size of 24 $\mu$m, which corresponds to $0\farcs254$ pixel$^{-1}$ 
along the spatial direction and 3.7 km~s$^{-1}$~pixel$^{-1}$ along 
the dispersion direction.
The slit width was $1\farcs0$, and the resultant instrumental FWHM
was 12\kms\ at the H$\alpha$ line.  

The echelle observations were made along the major axis of NGC\,1360
at the position angle (PA) of 27\arcdeg.
Three slit positions were used to cover the full extent of the 
major axis, as marked in Figure~1.
The integration time at each slit position was 600 s.
The angular resolution, determined by the seeing, was $\sim2\farcs0$. 
A mosaic of the H$\alpha$ echellograms is shown in Figure~2,
alongside an H$\alpha$ image oriented such that the major axis is 
parallel to the spatial direction of the echellogram.  

\section{Results}

\subsection{Morphology}

As shown in the H$\alpha$ image (Fig.\ 1), NGC\,1360 appears
elliptical with its major axis oriented along PA $\sim$ 30$^\circ$.
It lacks a sharp outer boundary because the surface 
brightness declines with distance until it drops below the 
background noise.
In order to estimate the size of NGC\,1360, we determined the 
background noise in the H$\alpha$ image and defined the nebular
edge to be at 3$\sigma$ above the background.
Using the threshold of 3$\sigma = 5.3 \times10^{-16}$ 
ergs~cm$^{-2}$~s$^{-1}$~arcsec$^{-2}$, NGC\,1360 extends 
$9' \times 7'$, as shown by the contour in Figure~1.  
At a distance of 350$^{+1,000}_{-150}$ pc as determined from the 
{\em Hipparcos} measurements \citep{Aetal98}, the linear size of 
NGC\,1360 would be $0.9 \times 0.7$ pc.  

While the overall surface brightness distribution of NGC\,1360
falls off radially, there are depressions of $\sim$20\% 
within 30$''$ of the central star and in the eastern hemisphere.
The spatial distribution of these surface brightness depressions 
do not appear to be random.  
The central depression extends into the eastern hemisphere and 
curves southward, while another distinct depression curves from 
the east to the north of the region surrounding the central star.
The cause of these depressions is either lower gas densities or a 
higher extinction.
If the latter is true, the non-random distribution of the 
depressions suggests that the obscuring material is internal 
to the nebula.
A quantitative analysis of the surface brightness of NGC\,1360
is presented in \S4.

At the northern end of the major axis there exist several nebular
knots extending from 240$''$ out to 350$''$ from the central star.  
Near the southern end of the major axis, there are also 
knots (marked by an arrow in Fig.~1), but they are fainter
and detected at just the 3$\sigma$ level.
The faint southern knots, the central star, and the bright
northern knots are not aligned along a straight line; instead,
they form a slightly curved arc with the star at the apex.
Such a configuration has been observed in other PNe and 
suggested nebular motion through the interstellar medium.

\subsection{Kinematics}

The echelle observations at the three consecutive slit 
positions have been mosaicked into an echellogram spanning the 
full major axis of NGC\,1360 and are presented in Figure 2.
This H$\alpha$ echellogram shows a position-velocity ellipse,
indicating an expanding shell in the nebula.
Furthermore, the position-velocity ellipse is tilted, with 
the southwestern and northeastern tips blue- and red-shifted,
respectively.
For an expanding ellipsoidal shell, the tilt of the
position-velocity ellipse results from the inclination of
the shell's polar axis with respect to the sky plane.
As the position-velocity tilt occurs along the major axis 
of NGC\,1360, its ellipsoidal shell must be prolate in shape,
with its southwestern tip oriented toward us and its 
northeastern tip away from us.

The velocity profile of the H$\alpha$ line near the central
star shows two broad components (see Fig.~3).
The approaching component has a heliocentric velocity of
$V_{\rm hel}$ = 25 km~s$^{-1}$ and a width of 
FWHM = 30 km~s$^{-1}$, while the receding component has
$V_{\rm hel}$ = 73 km~s$^{-1}$ and  FWHM = 37 km~s$^{-1}$.
These widths are much larger than the thermal width of the
H$\alpha$ line at 10$^4$ K, FWHM = 21.4 km~s$^{-1}$, indicating
the existence of significant velocity variations along the line 
of sight within the nebular shell.

The average velocity of the approaching and receding sides of 
the nebula, $V_{\rm hel}$ = 49 km~s$^{-1}$, is adopted as the 
systemic velocity of NGC\,1360.  
This result is in good agreement with the previous 
measurement of $V_{\rm hel}$ = +49.1 km~s$^{-1}$ by 
\citet{BS74} using a Fabry-Perot scanner for a 60$''$ 
aperture near the center of NGC\,1360.
The line-splitting at the center of a PN is usually adopted as 
twice its expansion velocity so that the apparent expansion 
velocity of NGC\,1360 would be 24 km~s$^{-1}$; however, 
as we discuss in \S4.1, this is an oversimplification of the 
expansion of NGC\,1360.

In the H$\alpha$ line, both the nebula and the knots are 
detected; however, in the [\ion{N}{2}] line, only the knots 
are detected.
This is illustrated by the H$\alpha$ and [\ion{N}{2}] lines of 
the northeastern end of NGC\,1360's major axis shown in Figure~4.
The low [\ion{N}{2}]/H$\alpha$ ratio of the nebula is expected 
because the central star of NGC\,1360 has an effective temperature
of 110,000 K \citep{Nap99} and the ionization stage of N in the
nebula is higher than N$^+$.
The high [\ion{N}{2}]/H$\alpha$ ratios of the knots suggest that
they may be FLIERS, i.e., fast, low-ionization emission regions
\citep{Betal93}.
Figure~4 shows that the innermost knots detected within the 
slit are at $\sim$260$''$ from the central star and have a 
radial velocity of $V_{\rm hel} \sim$100 km~s$^{-1}$ 
corresponding to a systemic velocity of $\sim$+50 km~s$^{-1}$.
The radial velocity of the knots increases outward by 
0.5 km~s$^{-1}$ arcsec$^{-1}$.
These knots clearly move faster than the shell expansion,
confirming their identification as FLIERs.

\section{Discussion}

\subsection{Spatio-Kinematic Structure of NGC\,1360}

The H$\alpha$ image and echelle spectra of NGC\,1360 can 
be used to construct spatio-kinematic models.
In contrast to the identifiable thin shells in the Owl 
Nebula (Paper I) and NGC\,7662 (Paper II) that can be 
modeled easily, the shell of NGC\,1360 has both large 
velocity widths and irregular brightness variations 
in the echelle line image, which require simplifying 
assumptions to make the modeling tractable.
We have assumed that NGC\,1360 consists of a series of 
concentric thin ellipsoidal shells and the shell 
expansion velocity increases outwards.
Consequently, the envelope of the position-velocity ellipse
(marked in Fig.~2a) corresponds to an outer thin ellipsoidal
shell that may be identified with the elliptical isophotes 
with a matching spatial extent in the direct image (marked in
Fig.~2b).
This outer thin shell can then be modeled using the 
methodology outlined by \citet{Getal03}, by assuming
that the shell expands homologously (i.e., a 
Hubble law).

We model this shell of NGC\,1360 as a prolate ellipsoidal
shell with its polar axis tilted with respect to the 
line-of-sight by an inclination angle $i$.
Its polar and equatorial radii are $r_{\rm p}$ and $r_{\rm e}$, 
and its polar and equatorial expansion velocities are 
$V_{\rm p}$ and $V_{\rm e}$, respectively.
For the assumed homologous expansion, $r_{\rm p}/r_{\rm e}$
is equal to $V_{\rm p}$/$V_{\rm e}$, and the kinematic age 
is simply $r_{\rm e}$/$V_{\rm e}$ or $r_{\rm p}$/$V_{\rm p}$.
A grid of models are constructed for a range of inclination 
angles $i$, axial ratios $r_{\rm p}$/$r_{\rm e}$, and
equatorial expansion velocities $V_{\rm e}$.
These models are then used to synthesize nebular isophotes
and to generate position-velocity ellipses for visual
comparisons with the observations.
Our best-fit model is marked in Figure 2, and has the 
following parameters:
$i = 60$\arcdeg, $V_{\rm e}$ = 35 km~s$^{-1}$ at 
$r_{\rm e}$ = 130\arcsec, 
$V_{\rm p}$ = 70 km~s$^{-1}$ at $r_{\rm p}$ = 260\arcsec,
and $t_{\rm kin}$ = 10,000 yr.
Note that the inclination angle and the axial ratio are
correlated so that lower values of $i$ yield higher ratios 
of $V_{\rm p}$/$V_{\rm e}$.
Despite this fact, the inclination angle and the axial ratio 
are both well constrained.
The former is within $\pm$10\arcdeg\ from the
optimal value, while the latter is within $\pm$0.25 
from $V_{\rm p}$/$V_{\rm e}$ = 2.0.
The expansion velocities and kinematic age, on the other 
hand, are not as well constrained.
For $i$ = 60\arcdeg, the uncertainty in $V_{\rm e}$ is
$\pm$10 km~s$^{-1}$, corresponding to about $\pm$30\%.

Bear in mind that this model describes only an outer
layer of the nebular shell of NGC\,1360.  
The bulk of the nebular material is interior to 
this layer and expands more slowly.
This densest shell material is responsible for the
peaks in the velocity profile.
For an inclination angle of 60\arcdeg, the velocity
split near the nebular center roughly corresponds to
twice the equatorial expansion velocity of the densest
nebular material in NGC\,1360, 24 km~s$^{-1}$ (see Fig.~3
and \S3.2).
However, the detailed geometry and kinematics of this
dense material cannot be more precisely determined 
because the peaks of the line profiles do not form
a well-defined position-velocity ellipse in the echelle
image.

The FLIERs in NGC\,1360 are projected along the
apparent major axis of the nebula; thus, it is
reasonable to assume that the FLIERs are distributed
along the polar axis with an inclination
angle of 60\arcdeg.
The apparent expansion velocities of the FLIERs, 
$\ge$50 km~s$^{-1}$, then correspond to deprojected 
expansion velocities of $\ge$100 km~s$^{-1}$.
These expansion velocities are typical of those 
observed in FLIERs \citep{BPI87}.

\subsection{Physical Structure of NGC\,1360}

We can use the spatio-kinematic model of NGC\,1360 deduced in 
\S4.1 along with the observed H$\alpha$ surface brightness 
distribution to study the density structure within this nebula.
The observed H$\alpha$ surface brightness profiles along several 
PAs are plotted in Figure 5. 
The depression in surface brightness seen in the eastern hemisphere
of NGC\,1360 is clearly visible in the profiles
along PAs between 105\arcdeg\ and 165\arcdeg.
In contrast, the surface brightness of the western hemisphere
of NGC\,1360 shows smoother profiles, without obvious 
``bumps" or ``dents." 
We have carried out the analysis of the density structure of 
NGC1360 using the sections of the nebula between PA 225\arcdeg\
and 345\arcdeg\ because their uniformity best represents the
overall distribution of material in the nebular shell.

The profiles of the western regions of NGC\,1360 in Figure 5 exhibit
a Gaussian shaped brightness distribution peaking at 
40\arcsec--60\arcsec\ and declining smoothly with increasing radius.
The smooth shapes of the surface brightness profiles indicate that the 
nebula does not contain any regions with sharp density or extinction
variations in the western hemisphere.
These surface brightness distributions, in conjunction with the
geometric information determined from the spatio-kinematic modeling 
in \S3.2, can be used to derive the nebular density distribution.
We have  modeled the surface brightness profile assuming a Gaussian 
density distribution and fitted this model to the observed H$\alpha$
surface brightness profiles between PA 225\arcdeg\ and 345\arcdeg.
The best-fit model peaks at 
35\arcsec\ with a FWHM of 60\arcsec\ along the equatorial plane
and peaks at 70\arcsec\ with a FWHM of 100\arcsec\ along the polar 
direction.
The best-fit models are overplotted in Figure 5: good agreements are
obtained from PA 225\arcdeg\ to 345\arcdeg.  
Using the flux calibrated surface brightness we further determined
that the peak density is $\sim$130 H-atom cm$^{-3}$, and that the 
density at $r_{\rm e}$ = 130\arcsec\ is $\sim$0.1 H-atom cm$^{-3}$.  
The low density at the outer edge of the nebula implies that it 
is beginning to merge into the interstellar medium.  
The total mass in the model, $\sim$0.4 M$_\odot$, is estimated assuming
that the surface brightness depressions in the eastern hemisphere are
caused by internal dust absorption.
If the lower surface brightness is caused by a lack of material in the
shell, the mass would be over-estimated, but this error is small as
compared to the uncertainty in mass due to the large error bar in the
distance to NGC\,1360.

\section{Summary and Conclusions}

We have obtained images and high-dispersion long-slit spectra 
of NGC\,1360 in order to study its spatio-kinematic structure.
We have modeled an outer layer of the shell as a prolate 
ellipsoid and find a major-to-minor axial ratio of 2:1 and a 
polar axis tilted by 60\arcdeg\ with respect to the line of sight.  
The kinematic age derived from this model is $\sim$10,000 yr.  

We have found that the observed morphology and surface 
brightness profiles of NGC\,1360 can be modeled well by
a thick shell with a Gaussian radial density profile.
This smooth density distribution implies that there is no
sharp inner edge as expected from compression by a fast 
stellar wind.
If a fast wind was present in the past and compressed the
AGB wind into a thin shell, that shell has since diffused.
The low nebular density, $\le$130 H-atom cm$^{-3}$, suggests 
that NGC\,1360 is an evolved PN; its low-density (0.1 H-atom 
cm$^{-3}$) envelope has begun to merge with the interstellar 
medium.  

The FLIERs observed in NGC\,1360 are intriguing.
The kinematic age of the FLIERs, based on their de-projected 
distances from the central star, $\ge$0.5 pc, and their
de-projected expansion velocities, $\ge100$ km~s$^{-1}$, 
is $\sim$5,000 yr.
This kinematic age is about 1/2 that of the PN itself.
NGC\,1360 is one of the very few cases of evolved PNe in 
which collimated outflows are observed \citep[e.g.,][]{Cetal99}.

\acknowledgments

We thank Jin-chung Hsu and Stefano Casertano for their help in 
addressing the warm pixel problem of the {\it HST} WFPC2 image.

\clearpage

\begin{figure}
\epsscale{0.90}
\caption{
An H$\alpha$ image of NGC\,1360 displayed with two different 
intensity contrasts to highlight the faint outer regions 
({\it left}) and the structure in the central regions ({\it right}).  
The contour in the left panel corresponds to the 3$\sigma$ 
level above the background.  
The slit positions used in the echelle spectroscopic
observations are marked on the image.  
The faint southern FLIERS are marked with arrows in both panels.
}
\end{figure}

\begin{figure}
\epsscale{0.90}
\caption{
({\it left}) 
H$\alpha$ echellogram along the major axis of NGC\,1360.  
The geocoronal H$\alpha$ line is marked.  
({\it right}) 
H$\alpha$ image of NGC\,1360 oriented to match its major axis 
with the echelle slit position for direct comparison.
The jagged contours overlaid on the echellogram and
direct image have matching spatial extent and correspond
to isophotes detected at 4$\sigma$ and 7$\sigma$ levels, 
respectively.
The smooth ellipses overlaid on the echellogram and 
direct image are based on our best-fit spatio-kinematic model.  
The bright FLIERs northeast to the nebula are visible in 
both panels.
}
\end{figure}

\begin{figure}
\epsscale{0.5}
\caption{
Velocity profile of the H$\alpha$ line within the central 
region of NGC\,1360.  
The position of the geocoronal H$\alpha$ line is marked.  
The systemic velocity adopted for NGC\,1360 is also marked.  
}
\end{figure}

\begin{figure}
\epsscale{0.666}
\caption{
An expanded view of the H$\alpha$ and [N~{\sc ii}] echellograms
at the northeast end of NGC\,1360 to show the FLIERs.
The spatial position is measured relative to the central star.
The H$\alpha$ line shows both the nebular shell and the 
FLIERs, while the [\ion{N}{2}] line shows only the FLIERs.
The FLIERs exhibit a velocity gradient along the slit 
visible in both lines.
}
\end{figure}

\begin{figure}
\epsscale{0.90}
\caption{
Surface brightness profiles (solid lines) of the H$\alpha$ 
line extracted at 30\arcdeg\ intervals.  
The peak at the zero position corresponds to the central star,
and the position angle is marked for the profile on either side
of the star.
The dashed curves are the profiles from our best-fit model.
}
\end{figure}


\begin{thebibliography}{}

\bibitem[Acker et al.(1998)]{Aetal98} 
Acker, A., Fresneau, A., Pottasch, S.~R., \& Jasniewicz, G.\ 1998, 
\aap, 337, 253 

\bibitem[Balick, Preston, \& Icke(1987)]{BPI87} 
Balick, B., Preston, H.\ L., \& Icke, V.\ 1987, \aj, 94, 1641 

\bibitem[Balick et al.(1993)]{Betal93} Balick, B., Rugers, M., 
Terzian, Y., \& Chengalur, J.~N.\ 1993, \apj, 411, 778 

\bibitem[Bohuski \& Smith(1974)]{BS74} 
Bohuski, T.~J.~\& Smith, M.~G.\ 1974, \apj, 193, 197 

\bibitem[Corradi et al.(1999)]{Cetal99} 
Corradi, R.\ L.\ M., Perinotto, M., Villaver, E., Mampaso, A., 
\& Gon{\c c}alves, D.\ R.\ 1999, \apj, 523, 721 

\bibitem[Frank \& Mellema(1994)]{FM94} Frank, A.~\& Mellema, 
 G.\ 1994, \apj, 430, 800 

\bibitem[Guerrero et al.(2003)]{Getal03}
Guerrero, M.~A., Chu, Y.-H., Manchado, A., \& Kwitter, K.~B.\ 2003, 
\aj, 125, 3213 (Paper I)

\bibitem[Guerrero et al.(2004)]{Getal04}
Guerrero, M.~A., Jaxon, E.~G., \& Chu, Y.-H.\ 2004, AJ, in press, October 
issue (Paper II)

\bibitem[Kwok et al.(1978)Kwok, Purton, \& Fitzgerald]{KPF78} Kwok, S.,
Purton, C.~R., \& Fitzgerald, P.~M.\ 1978, \apj, 219, L125

\bibitem[Napiwotzki(1999)]{Nap99} Napiwotzki, R.\ 1999, \aap, 
350, 101 

\end{thebibliography}
\end{document}